\documentclass[twocolumn,amssymb,amsmath,
nofootinbib,tightenlines,showpacs,floatfix,
superscriptaddress,aps,pra]{revtex4-1}
\usepackage{amsfonts}
\usepackage{bm}%
\usepackage[colorlinks=true,linkcolor=blue,filecolor=blue,menucolor=yellow,urlcolor=blue,
citecolor=blue,anchorcolor=blue]{hyperref}
\usepackage{graphicx}
\usepackage{dcolumn}
\usepackage{rotating}

\begin{document}
\title{Superfluid Field response to Edge dislocation motion}

\author{Abdul N. Malmi-Kakkada}
\email{naseermk@utexas.edu}
\affiliation{School of Physics and Astronomy, University of Minnesota,
Minneapolis, Minnesota 55455}
\affiliation{Department of Chemistry, University of Texas, Austin, Texas, 78712}

\author{Oriol T. Valls}
\email{otvalls@umn.edu}
\altaffiliation{Also at Minnesota Supercomputer Institute, University of Minnesota,
Minneapolis, Minnesota 55455}
\affiliation{School of Physics and Astronomy, University of Minnesota, 
Minneapolis, Minnesota 55455}

\author{Chandan Dasgupta }
\email{cdgupta@iisc.ernet.in}
\altaffiliation{Also at Condensed Matter Theory Unit, Jawaharlal Nehru Centre
for Advanced Scientific Research, Bangalore 560064, India}
\affiliation{Centre for Condensed Matter Theory, Department of Physics, 
Indian Institute  of Science, Bangalore 560012, India}

\date{\today}
\begin{abstract}

We study
the dynamic response of
a superfluid field  to a moving edge dislocation line 
to which the field is minimally coupled.
We use a dissipative Gross-Pitaevskii equation, and determine
the initial conditions by solving the equilibrium
version of the model.
We consider the subsequent time
evolution of the field for both glide and climb dislocation motion and
analyze the results for a range of  values of 
the constant speed $V_D$ of the moving dislocation.
We find that the type of motion of the dislocation line is 
very important in determining the time evolution of the
superfluid field distribution associated
with it. Climb motion of the dislocation line 
induces increasing  asymmetry, as function of time, in the field
profile, with part of the probability being, as it were,
left behind.  On the other hand, glide motion 
has no effect on the symmetry properties of the superfluid field distribution.
Damping of the superfluid field 
due to excitations associated with the moving dislocation line occurs
in both cases.

\end{abstract}


\maketitle

\section{Introduction} 
\label{Introgpe} 

Supersolids represent an exotic state of quantum matter in which two kinds of order exist
simultaneously: crystalline order associated with the breaking of translational symmetry and
superfluid order associated with the breaking of the symmetry under a global rotation
of the quantum mechanical phase. The possibility of occurrence of a supersolid phase in solid $^4$He
was pointed out~\cite{helium1,helium2,helium3} many years ago. More recently, supersolid phases have been 
realized in experiments on ultracold atomic systems. 
The first experimental observations of a supersolid phase~\cite{ss1,ss2,ss3} involved the self-organization of a Bose-Einstein condensate (BEC) in which the discrete translational
symmetry of a preimposed lattice structure was spontaneously broken. 
Recent experiments~\cite{ss4,ss5} have demonstated the occurrence of supersolid
phases in which the continuous translational symmetry of space is spontaneously broken. One of these 
experiments~\cite{ss4} involves a BEC with spin-orbit coupling that exhibits a supersolid stripe phase
with density modulation in one direction. The other experiment~\cite{ss5} 
involves a BEC dispersively coupled to
two optical cavities.  Many other proposals for the occurrence of supersolid phases in ultracold atomic systems
exist in the literature~\cite{ssrev,ssa,ssb,ssc} and it is expected that some of these proposals will be realized in experiments in the
near future. Therefore, studies of various physical  properties of quantum solids in the presence
of superfluidity are of much current interest.

Topological defects of crystalline solids are dislocations which form a network of lines in three dimensional systems. 
There exists a vast literature~\cite{hull1} on the properties of dislocations in conventional solids. Dislocations play a very 
important role in the mechanical response of crystalline solids. When a crystalline
solid is subjected to an external stress, the response of the solid is determined to a large extent by the motion of
dislocations induced by the stress. The properties of dislocations in supersolids is relatively less understood. The motion of a dislocation
line in a supersolid is more complicated than that in a conventional crystal because of the presence of superfluidity.
The motion of a dislocation line in a supersolid affects the superfluidity in its vicinity because the superfluid order parameter is
coupled to the strain field of the dislocation line. This coupling also changes parameters such as the mobility associated
with the motion of dislocation lines. This interplay between the motion of a dislocation line and superfluidity in its
neighborhood is the subject of our study. Dislocations in the vortex lattice in a rotating BEC~\cite{vortex_lattice}
provide another example of a cold matter system in which this interplay between crystal defects and 
superfluidity plays an important role. 

This subject is also important for understanding the low-temperature properties of solid $^4$He. 
Interest in the old predictions~\cite{helium1,helium2,helium3}
of occurrence of supersolidity in $^4$He was renewed when Kim and Chan~\cite{kimi} observed 
a period drop in torsional oscillator (TO) experiments with solid $^4$He and interpreted
the observation as evidence for the occurrence of a supersolid phase in this
system at sufficiently low temperatures. 
Evidence that structural disorder present in samples of solid $^4$He could play an important role became apparent early on~\cite{rittner,rittner1,sample,balibarh}: 
results of TO experiments were found to depend strongly on sample preparation methods, 
and annealing the sample was found to substantially reduce 
the TO period drop. 
Subsequently, it was found~\cite{day,day1} that an elastic anomaly with a jump in 
the shear modulus  occurs in solid $^4$He at a temperature close to that of the
TO period drop. It was soon realized~\cite{iwasa,maris} that this elastic 
anomaly can account for the TO period change.
At present, the emerging consensus~\cite{reppy1,kim2,kim1} seems to be that the anomalous low-temperature properties of solid $^4$He can be understood entirely in terms of the stiffening of the solid, without having to invoke the occurrence of superfluidity. The elastic anomaly is attributed to the pinning 
of dislocation lines by $^3$He impurities which prevent the dislocation lines from
gliding along basal planes in solid $^4$He. This description accounts for several
experimentally observed features~\cite{haziot1,haziot2,fefferman} in the elastic 
properties of solid $^4$He at low temperatures.

There are, however, several experimental 
and theoretical results that suggest that the occurrence of superfluidity in the vicinity of defects such as dislocation lines 
and grain boundaries in solid  $^4$He may play
an important role in the low-temperature properties of solid $^4$He. Experimental
observations~\cite{mass-flux,mass-flux1} 
of mass flow through solid $^4$He have been attributed to flow of atoms 
through 
superfluid dislocation cores. It has been suggested~\cite{superclimb} that mass flow through superfluid cores of edge dislocations 
 can lead to ``superclimb'' that would provide 
  an explanation of the large isochoric compressibility observed in Ref.~\cite{mass-flux}. Results of experiments~\cite{rotation} on the effects of dc rotation on the TO period drop 
also suggest the occurrence of superfluidity. There are reports~\cite{souris,rojas1,rojas2} of the occurrence of an elastic anomaly in ultra-pure samples of solid $^4$He in which the spacing between $^3$He impurities is expected to be of the order of or larger than the size of the sample. The stiffening of the solid in these samples can not be attributed to the pinning of dislocation lines by $^3$He impurities. 
It has been shown recently~\cite{jltp_paper} that the onset of
superfluidity in and around the cores of dislocation lines can lead to an increase of
the shear modulus of the solid by decreasing the mobility of the dislocation lines.
Quantum Monte Carlo calculations~\cite{svist,pollet,boninsegni} have 
shown that superfluidity can occur in the vicinity of structural defects such as
dislocations and grain boundaries in solid $^4$He. Theoretical studies~\cite{toner,yoo}
indicate that a generic coupling between the superfluid field and the elastic strain field 
associated with a dislocation line in a phenomenological 
Landau theory of superfluidity leads to superfluidity in the vicinity of
a stationary edge dislocation line. Theoretical studies~\cite{shev,toner,yoo,cdotv} 
have also shown that bulk superfluidity can occur in solid  $^4$He from superfluidity along a network of crystal defects. It is clear from these results that studies of
superfluidity near dislocation lines in solid  $^4$He are important, even if bulk superfluidity does not occur in this system.

\begin{figure}
\includegraphics[width=0.45\textwidth] {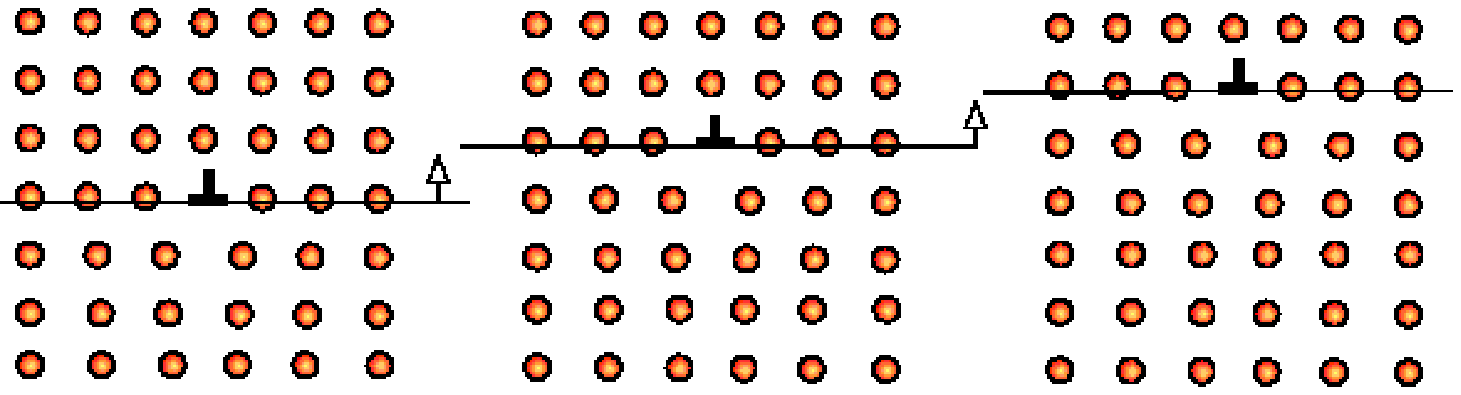}
\par
\vspace{1.0 cm} 
\includegraphics[width=0.45\textwidth] {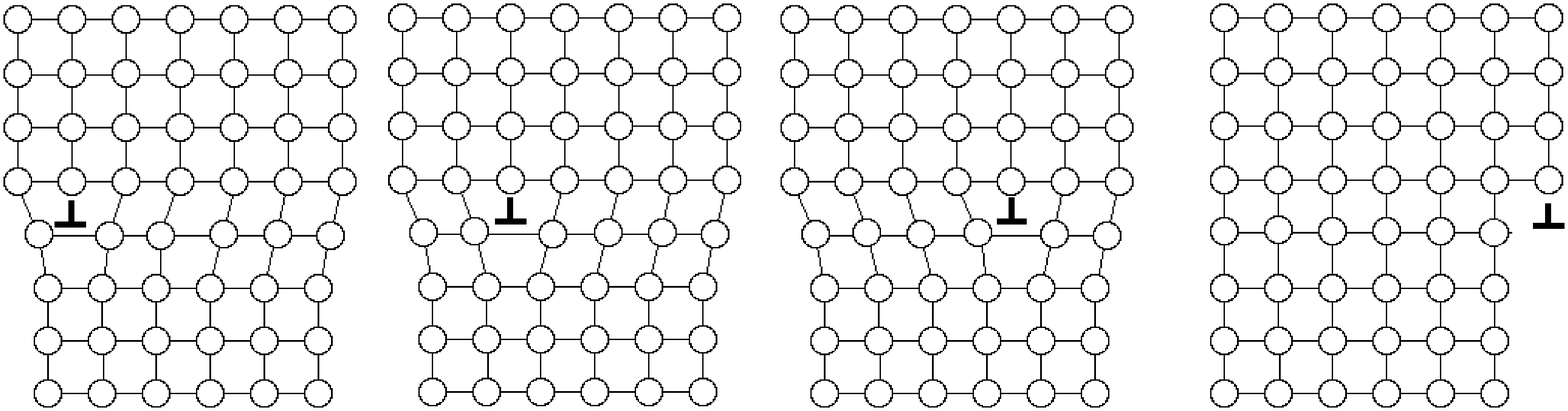}
\caption{ Top panel: an edge dislocation is illustrated executing 
climb motion. The arrow indicates the direction of climb.
The bottom panel illustrates the glide motion of an edge dislocation. 
}
\label{fig0climbglide}
\end{figure}

In this paper, we make the assumption 
that superfluidity occurs near a dislocation line in a quantum crystal and examine the effect of motion of the dislocation line on superfluidity in its vicinity. Previous studies~\cite{pollet,boninsegni,shev,toner,yoo} of superfluidity near a 
dislocation line focused on the case
where it 
is quenched or stationary. Many physical effects, such as
the ``giant plasticity'' of solid $^4$He, attributed to nearly free gliding motion 
of edge dislocations along the basal planes, involve moving dislocation lines.
Therefore, it is important to understand the
effects of the motion of a dislocation line on the superfluidity near its core. 
Dislocation lines are dynamic objects that execute a variety of motions. 
Dislocation line segments can undergo roughening~\cite{roughening}
and can execute two basic types of motion in response to an applied stress: 
climb or glide motion. These two different types 
of motion are illustrated in Fig.~\ref{fig0climbglide}. 
When a dislocation line moves along the surface that contains both itself
and the Burgers vector associated with it, the motion is called glide. 
Movement out of the glide surface in a direction perpendicular to the Burgers
vector is referred to as climb. 
Glide and ``superclimb'' i.e. climb assisted by superfluidity 
in the dislocation cores in solid $^4$He were studied previously~\cite{aleinikava}
in the context of elastic effects such as dislocation line tension and 
compressibility. Dislocation lines can glide freely along basal planes
in solid $^4$He at relatively high speeds compared to other crystals. 
This is thought to be a quantum effect which causes the Peierls barrier to 
dislocation motion to be negligible~\cite{haziot1}. 

Our main interest in the present study is to investigate 
whether the motion of a dislocation line 
modifies (e.g. enhances, suppresses, or distorts)  
the associated superfluid field. 
It has been suggested~\cite{balibar_nature} that transverse motion of a dislocation
line reduces the degree of superfluid ordering in its vicinity.
A study~\cite{ising} of a toy model of Ising spins residing on the
links of a random network shows that motion of the links leads to a reduction in the
ordering temperature of the spins. A similar effect is expected for superfluidity along
a network of dislocation lines. Thus, we study here 
the effect of the dynamics of a dislocation line on the superfluid field
at the microscopic quantum level.
We have analyzed the 
response of the superfluid field near an edge dislocation 
line, assumed to be driven at constant velocity $\vec{V}_D$, 
for both climb and glide motion. 
Focusing on small length and time scales, we have taken into account fluctuations in both the amplitude and the phase of the superfluid order parameter. 
Fluctuations in the amplitude were not included 
in the coarse-grained models studied earlier, because the amplitude is  not a 
hydrodynamic variable.

A mathematical framework that has been used extensively for describing superfluidity 
in $^4$He is the Gross-Pitaevskii 
equation (GPE)~\cite{gross, pita1}. The GPE, also referred to as the
nonlinear Schr\"{o}dinger equation, has been quite successful in 
helping understand the equilibrium and dynamic 
behavior of low-temperature superfluids and Bose-Einstein condensates (BEC)~\cite{review2}. 
However, the GPE does not provide a description of damping 
and can only be used to study dissipationless fields.
As we are interested in exploring what quantum models predict for the damping
of the superfluid field near a dislocation line due to its motion, a method to include 
dissipation in the GPE is necessary. 
With this purpose in mind, an approach 
similar to that used in Ref.~\cite{pita} to  
study damping of superfluidity near the $\lambda$ point is used in our study. 
The modification of the GPE in order to capture the effects
associated with damping is referred to as the Dissipative Gross-Pitaevskii equation (DGPE)~\cite{burnett}. 
Based on the DGPE formalism and a well-studied model~\cite{toner,yoo} for the coupling
of the superfluid field with the strain field 
of a dislocation line, we present in this work a study of how the excitations 
associated with a moving dislocation line in solid $^4$He
affects the superfluid field near it. 
Both GPE and DGPE approaches have been
used previously~\cite{kumini2014,cidrim2016} to study the effects of moving a line object 
in a superfluid. Our work is similar in spirit to these earlier studies.

We find that the motion of the dislocation line 
plays an important role in determining the  
superfluid field distribution near it. During climb motion, 
a part of the superfluid field 
associated with a stationary dislocation line is ``left behind." 
Climb induces more asymmetry in the distribution 
of the superfluid field near the dislocation line. 
No effect on the symmetry properties of the superfluid 
wavefunction is observed for glide motion. 
Decay of the superfluid field amplitude during climb and glide is observed, but the
magnitude of the decay is very 
small for experimentally realistic values of the dissipation parameter. 

The rest of this paper is organized as follows: Introduction of the DGPE formalism 
and details of the parameter values used are presented in Section~\ref{methodgpe}.
We also provide in this Section the details of how the elastic strain field due to the
dislocation is coupled to the superfluid order parameter. Results of our study of  
the effects of the motion of a single edge dislocation line on the superfluid field are presented in Section~\ref{resultgpe}. The main conclusions of our study are 
summarized in Section~\ref{conclusion}.

\section{Methods} 
\label{methodgpe}

\subsubsection{Dissipative Gross-Pitaevskii equation}
We consider in this study
a single long, straight edge dislocation line running along the $z$ axis. 
The Burgers vector 
for the edge dislocation is taken to be in the $x$-direction. 
The dislocation line is assumed to be long and 
straight so that one can 
 neglect edge effects and define the problem 
in the 2D $x - y$ plane orthogonal to it. 
The standard GPE which describes the motion of a field, $\psi$, 
is of the form 
\begin{equation}
\label{gpeeq}
i\hbar\frac{\partial\psi}{\partial t} = \frac{-\hbar^2}{2m}\nabla_{x,y}^2\psi + v(x,y;t)\psi + g|\psi|^2\psi, 
\end{equation}
where $\nabla_{x,y}^2 = \frac{\partial^2}{\partial x^2} +  \frac{\partial^2}{\partial y^2}$, 
$m$ is the mass of an atom, $v(x,y;t)$ is the potential
(the time dependence arises from the dislocation
motion) 
and $g$ is the superfluid interaction parameter. On the right hand side, 
the first two terms are the kinetic and potential energy 
and the third, nonlinear term describes the interaction energy 
between superfluid atoms. This interaction is repulsive, 
$g > 0$. It is given 
by~\cite{hadzi,ssrev}
\begin{equation}
\label{g2dgpe}
g = \frac{4\pi\hbar^2 a_s N}{m L}.
\end{equation}
Here $a_s$ is the microscopic s-wave 
scattering length, 
$N$ is the number of superfluid atoms and $L$ is the size
of the trap. 

In the problem under study, 
the complex field $\psi$ is the superfluid wavefunction, and 
the coupling between $\psi$ and the dislocation strain potential is introduced
via the term $v(x,y;t)\psi$~\cite{toner}. 
For an edge dislocation along the $z$-axis 
the 
strain potential is of the form~\cite{toner}
\begin{equation}
\label{pogpe}
v(x,y) = \frac{A}{\sqrt{x^2+y^2}}\cos\phi, 
\end{equation}
where $\phi = \arctan(x/y)$ is an azimuthal angle 
defined in the $x-y$ plane with respect to the $y$ axis. 
The parameter $A$, a positive quantity, denotes the strength 
of the dislocation potential 
and depends on the lattice and elastic
constants of the solid. For $A > 0$, this potential 
is attractive for $y < 0$ thereby allowing for bound states. 
For $y > 0$ the potential is repulsive. The potential is
symmetric along the $x$ axis (i.e. along the direction of the Burgers vector). 
These characteristics of the potential 
should be reflected on the wavefunction $\psi$ as well. 

The solution of the non-linear equation noted above 
is complicated due to the non-central nature of the potential~\cite{yoo1}. 
The equilibrium steady state of the superfluid field 
at very low temperatures, $T \rightarrow$ 0, 
is described by the time independent GPE
\begin{equation}
\label{gpeti}
-\frac{\hbar^2}{2m}\nabla_{x,y}^2\psi + v(x,y)\psi + g|\psi|^2\psi = \mu\psi, 
\end{equation}
where $\mu$ is the chemical potential. For the equilibrium solution, 
the wavefunction is normalized according to
\begin{equation}
\label{normgpe}
\mathcal{N} = \int^{+\infty}_{-\infty} \int^{+\infty}_{-\infty} |\psi|^2\,dx dy = 1. 
\end{equation}

The standard GPE (Eq.~(\ref{gpeeq}) above) contain no dissipative terms. 
The motion of the dislocation line is actually 
dissipative 
as a result of the various damping mechanisms within the crystal 
mentioned below. 
To account for dissipation  in the GP formalism 
we introduce  into the GPE  
a dimensionless damping factor $\gamma$, as in Ref.~\cite{burnett}.
The resulting
dissipative GPE (DGPE) 
is of the form
\begin{equation}
\label{dgpeeq}
i\hbar\frac{\partial\psi}{\partial t} = 
(1-i\gamma)[-\frac{\hbar^2}{2m}\nabla_{x,y}^2\psi + v(x,y)\psi + g|\psi|^2\psi -
\mu\psi],
\end{equation}
where the positive damping factor $\gamma$ is phenomenologically introduced 
in a way similar to that in Ref.~\cite{pita}. 
The right hand side terms in the square bracket 
represent the change from the equilibrium state of the 
superfluid wavefunction due to dynamics,  
in our case the moving dislocation line. 
The damping factor $\gamma$ 
is inversely proportional 
to a relaxation time and due to it neither the 
energy nor $\mathcal{N}$ are 
conserved in Eq.~(\ref{dgpeeq}).
In the original study by Pitaevskii~\cite{pita}, $\gamma$ was 
expressed in terms of the second viscosity coefficients of 
superfluid Helium. 
A similar equation with the factor 
$(1-i\gamma)$ was used in the study of soliton
decay and damping of vortices~\cite{kasem,parker}.

The dynamics of the $\psi$ field and its damping due to 
elementary excitations from a moving dislocation line 
can now be studied within the framework of Eq.~(\ref{dgpeeq}). 
By numerically solving the two-dimensional (2D)
time dependent DGPE with a moving dislocation line
(either climb or glide motion), 
the response of the superfluid order parameter
$\psi$ can now be evaluated. 
We consider the scenario where the dislocation line executes glide or climb 
at a constant velocity $\vec{V}_D$ due to external forces. 

Prior to presenting the details of the numerical calculation,
we need to give  
an overview of the  units used. 
It is convenient to 
rescale the length and time in terms of natural units. 
We choose for our unit of length the elastic correlation length $\xi_{el}$
defined by equating the kinetic energy 
of the superfluid 
to the potential energy due to the 
dislocation line, ${\hbar^2}/{2m\xi_{el}^2} = {A}/{\xi_{el}}$. 
Similarly, we rescale time by  
the characteristic frequency $\omega_{el} \equiv {\hbar}/{2m\xi_{el}^2}$. 
Rescaling the wave function, the cartesian co-ordinates 
$x,y$ as defined above, 
and the time via 
the definitions $\bar{t} \equiv \omega_{el}t$, $\bar{\psi} \equiv \psi\xi_{el}$,
 $\bar{x} \equiv x/\xi_{el}$ (similarly for $\bar{y}$), 
$\bar{v} \equiv v/\hbar\omega_{el}$, $\bar{g} |\bar{\psi}|^2\equiv g |\psi|^2/\hbar\omega_{el}$ 
and $\bar{\mu} \equiv \mu/\hbar\omega_{el}$ one obtains
\begin{equation}
\label{dgpeeq1}
i\frac{\partial\bar{\psi}}{\partial \bar{t}} = (1-i\gamma)
[-\bar{\nabla}_{\bar{x},\bar{y}}^2 + \bar{v}(\bar{x},\bar{y};\bar{t}) + 
\bar{g}|\bar{\psi}|^2 - \bar{\mu}]\bar{\psi}. 
\end{equation} 
The coefficient of the non-linear term 
is $\bar{g} \equiv {2mg}/{\hbar^2}$ and the strength of the dislocation 
potential, $A$,  
is rescaled such that $\bar{A} = A/\hbar\omega_{el}\xi_{el} = 1$ 
consistent with the definition 
of $\xi_{el}$ and $\omega_{el}$. 

\subsubsection{Numerical parameters and  initial condition}

We now discuss  
the numerical values of the parameters used in solving Eq.~(\ref{dgpeeq1}). 
The time dependent strain potential $v(x,y;t)$ in the DGPE 
depends upon whether the dislocation 
line is climbing or gliding.   
For climb 
motion, along the positive $y$-axis (perpendicular to the Burgers vector),
the dislocation potential depends on the speed $V_D$  
via
\begin{equation}
\label{tpogpe}
v(x,y;t) = \frac{A}{\sqrt{x^2+(y-V_{D}t)^2}}\cos\phi. 
\end{equation}
When the dislocation is caused to move in the 
direction of the Burgers vector along the $x$ axis,  
i.e. with the corresponding potential being
\begin{equation}
\label{tpogpe1}
v(x,y;t) = \frac{A}{\sqrt{(x-V_{D}t)^2+ y^2}}\cos\phi 
\end{equation}
it executes glide motion. 
Climb and glide motion of the dislocation line are 
considered separately in this study. 

The magnitude of the 
climb and glide velocity in classical crystals is expected to be small especially at low temperatures. 
In a quantum crystal such as solid $^4$He, however, the possibility of superclimb
and glide assisted by superfluidity~\cite{aleinikava}
requires one to include larger values of the velocity. 
Glide velocities up to 0.01 $m/s$ 
are considered in an experimental
study~\cite{haziot2} of dislocation velocities in solid $^4$He.
We take $V_D$ near its upper range to be better able 
to numerically observe its effects. 
To estimate the order of magnitude of $\xi_{el}$, the
strength of the dislocation potential $A$ (see Eq.(\ref{pogpe})) a 
characteristic 
quantity with dimensions of 
energy$\times$length, is needed. The magnitude of the parameter $A$ depends on 
the energy per unit length 
of an edge dislocation line, $E_{el} = Gb^{2}$, where $G$ is the 
shear modulus of the material 
and $b$ the magnitude of the Burgers vector~\cite{hull1}. 
For $A = E_{el} b \xi_{el}$ and using the definition of 
$\xi_{el} = {\hbar^2}/{2mA}$,  
for $G \sim 60~bar$ and $b \sim 10^{-10}~m$ appropriate
to solid $^{4}$He~\cite{fefferman}, 
one obtains $\xi_{el} \sim 10^{-9}m$.
This turns out to be roughly of the same order as the healing
length\cite{pethick} 
$\xi_{SF}$. 
Using the definition of   
$\omega_{el} \equiv \hbar/2m\xi_{el}^2$,
we obtain $\sim 10^{10}Hz$. 
Hence, $\xi_{el}\omega_{el} \sim 10~m/s$. 
The magnitudes 
 of the quantities $\xi_{el}$ and $\omega_{el}$ set 
the scale for length and time dimensions in the simulation, respectively. 
The natural units for $V_D$ are $~\xi_{el}\omega_{el}$. 
Dimensionless values for the magnitude of $V_D$ ranging from $5\times10^{-4}$ to $1.5\times10^{-3}$ 
(i.e. between 0.005$~m/s$ and 0.015$~m/s$, consistent with 
the experimental\cite{haziot2} range) 
are used for both climb and glide motion in our computations. 
These are much smaller than the speed of sound, which
is on the order of $10^2 m/s$.  

The strength of the interaction coefficient can be re-written as 
$\bar{g} \equiv 8\pi a_s \rho_{2D} \xi_{el}$  given $\bar{g} \equiv 2mg/\hbar^2$
and Eq.~(\ref{g2dgpe}). 
Here, the number density of superfluid atoms in 2D is $\rho_{2D} = N/\xi_{el}^2$.
The atomic number density 
of solid $^4$He is $\rho_{3D} = 10^{28}/m^{3}$~\cite{boningpe}. 
For the spacing between atomic planes at 
$\sim 3\times10^{-10}$m~\cite{aziz}, the number density in 2D (i.e. per atomic plane)
is $3\times10^{18}/m^{2}$. The scattering length, $a_s$, for $^4$He atoms is 
$\simeq 10^{-10}m$~\cite{pethick}. 
The strength of the nonlinear interaction parameter thus obtained is 
$\bar{g} \sim 7.5$. 
Since we focus on the superfluid condensate 
density near a dislocation line, the above 
value for $\rho_{2D}$ would greatly overestimate 
the condensate 
density. Assuming that a small percentage of atoms of order $1\%$ 
condense into the superfluid state~\cite{boninsegni,Saslow} near the dislocations, a smaller
value for $\bar{g} \sim 0.075$ is obtained. This is the value of $\bar{g}$ we use throughout this study. 

In order to solve Eq.~(\ref{dgpeeq1}), the value of the chemical potential $\bar{\mu}$ 
is needed. To obtain $\bar{\mu}$, the steady state GPE (see Eq.~(\ref{gpeti})) is numerically
solved using a relaxation method under the condition
that $\psi$ satisfies Eq.~(\ref{normgpe}). 
The accuracy of this method was tested using the 
two dimensional Coulomb potential, the solutions 
of which are well known~\cite{yoo1}. 
Length and time units were properly rescaled in terms of the units in 
Ref.~\cite{yoo1}
for these purposes. 
Using this procedure, the value 
$\bar{\mu} = -0.13$ was obtained for our system, 
consistent with other 
calculations~\cite{yoo1} of the same parameter for 
a two dimensional Schr\"{o}dinger equation with 
a non-moving dislocation line potential. 

The value of the dimensionless damping parameter $\gamma$ is also needed in order to solve 
the DGPE. In Ref.~\cite{burnett}, the magnitude of $\gamma$  
was found to depend 
 on the rate at which thermal particles above the Bose-Einstein condensate band
enter the condensate. This rate, compared to the relevant trap frequency,
sets the order of magnitude of $\gamma$. 
Using a similar approach, comparing the energy
dissipated by a moving dislocation line to the energy scale 
$\hbar\omega_{el}$, an estimate for $\gamma$ appropriate 
for the problem under consideration can be obtained. The energy dissipated during 
dislocation motion is roughly $F_{D}L_{D}b$ 
where $F_D$ is the force per unit length applied on a dislocation, 
$L_D$ is the typical length of a dislocation line and $b$ the magnitude of Burgers vector. 
The orders of magnitude of these quantities for a dislocation line
in solid $^4$He were obtained from Ref.~\cite{fefferman} 
where $F_D \sim 10^{-12} - 10^{-13} N/m$ and $L _D\sim 10^{-4} m$.  
The value of the parameter $\gamma \sim F_{D}L_{D}b/\hbar \omega_{el}$ thus obtained is $\sim 10^{-3}$.  
In our calculations, we set $\gamma=10^{-3}$, unless stated otherwise.

Next, we need to discuss the
 initial conditions chosen  and the numerical method 
used in our  computations. 
The equilibrium
solution obtained from the time independent GPE (Eq.~(\ref{gpeti})) is set
as the initial condition for $\psi$ in solving the time dependent 
Eq.~(\ref{dgpeeq1}). 
At, $\bar{t}=0$, the dislocation line is stationary and 
the superfluid distribution around it corresponds to the equilibrium case. 
As the dislocation line starts to move, the superfluid field $\psi$
near it reacts. The response of the superfluid field is 
studied for both glide and climb motion separately. 
Eq.~(\ref{dgpeeq1}) is solved using a split-step Crank-Nicolson method 
as presented in Ref.~\cite{adhikari}.
For the simulation, a 1200 $\times$ 1200 square grid system with the 
size of each grid being $0.05~\xi_{el}$ is used.  
We use fixed boundary conditions with $\psi \equiv 0$ 
at the boundaries of the computational grid.
A time step of $\delta\bar{t} = 0.01$ turns
out to be adequate. A small cutoff of 0.005 
for $\bar{x}$ and $\bar{y}$ 
is used in order to avoid the singularity associated with 
the dislocation potential at the origin. 
To avoid the possibility of an abrupt reaction of the superfluid field 
when $\vec{V}_D$ is switched on suddenly, we turn on the velocity of the 
dislocation line gradually over a time $\bar{t}_0$, short 
compared with the maximum simulation time, starting at zero
and ending at the desired value of $\vec{V}_D.$ 
In  the course of this initialization, the nonlinearity 
parameter $\bar{A}$ is slowly incremented to its value.
The value  $\bar{t}_0~\sim~100$ is used. The results have been verified to be independent of the small 
cutoff and insensitive to the precise value 
of $\bar{t}_0$. 

The scenarios considered here can easily be
related to experimental situations. 
Applying a stress on $^4$He crystal causes the dislocation
line to move. 
The contribution of factors such as 
thermal phonons or other impurities present in the crystal to  
damping of dislocation motion was discussed 
in Ref.~\cite{fefferman}. 
The parameter $\gamma$ takes into account
the effect on the superfluid field due to such excitations, as they may be 
induced by the dissipative motion of a dislocation line.
In the results presented below, we investigate how climb and glide motion
affects the superfluid field in its vicinity. 

\section{Results}
\label{resultgpe}
In this Section, we present the results of our DGPE simulation coupling 
a moving edge dislocation line to superfluidity. 
We analyze the effect of the motion of a dislocation line on 
a superfluid field assumed to be 
associated with its core. 
The first part of this section deals with climb and the latter part 
with glide motion of the dislocation line. 
As explained above, all lengths will be given in units of $\xi_{el}$, 
time in units of $\omega_{el}$ and velocity 
in terms of $\xi_{el}\omega_{el}$. 

\begin{figure}
\includegraphics[width=0.45\textwidth] {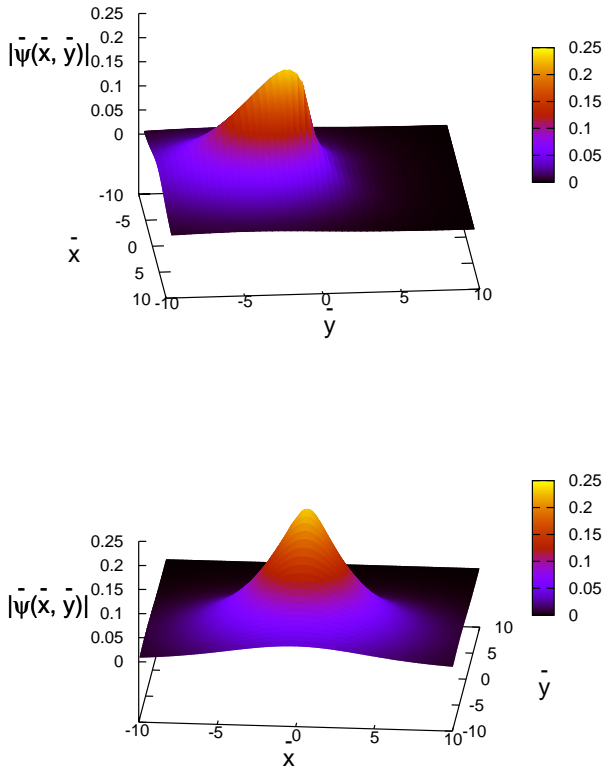} 
\caption{The absolute value of
the equilibrium  $|\bar{\psi}(\bar{x},\bar{y})|$ of the 
superfluid field. A broad maximum is seen in the attractive
region of the dislocation strain potential.
The two plots correspond to different
visual  orientations (see text).  Results were obtained by solving 
Eq.~(\ref{gpeti}). 
} 
\label{fig0gpe}
\end{figure}

To obtain the initial condition, the time independent GPE (Eq.~(\ref{gpeti})) is solved  
to get the equilibrium solution
for the superfluid field $|\bar{\psi}|$ near an edge dislocation line, using the stationary potential 
as given in Eq.~(\ref{pogpe}). Results are shown in Fig.~\ref{fig0gpe}.
The absolute value of the 
dimensionless equilibrium wave function, $|\bar{\psi}|$, near an edge dislocation line,  
is plotted there. The results
are given in the form of 
3D plots of the
superfluid distribution. Two different
viewing orientations are shown in that figure, 
specifically, views 
along the $\bar{y}$ axis and $\bar{x}$ axis are presented in the top 
and bottom panels respectively.  
It can be seen that a bound state of the superfluid field
forms in the attractive part of the dislocation potential 
(in the $\bar{y} < 0 $ region). The 
dislocation potential is 
symmetric along the $\bar{x}$ axis with respect to the origin, and 
asymmetric along the $\bar{y}$ axis. 
The symmetry characteristics of the potential can be seen 
reflected in $|\bar{\psi}|$: 
an asymmetric accumulation 
of the superfluid field in the region $\bar{y} < 0$ can be observed. 

\begin{figure}
\includegraphics[width=0.45\textwidth] {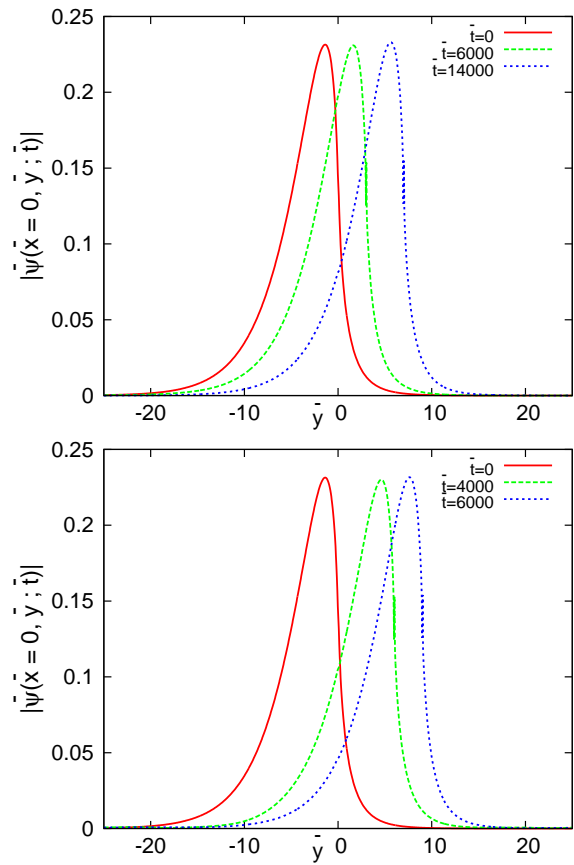} 
\caption{Plot of the absolute value of the 
wavefunction, $|\bar{\psi}(\bar{x}=0,\bar{y};\bar{t})|$ 
at different times for a climbing edge dislocation line. The top panel 
corresponds to $V_D = 5\times10^{-4}$ and 
the bottom panel to $V_D = 1.5\times10^{-3}$. The times are indicated in the
legend.} 
\label{fig1gpe}
\end{figure}

To study the effect of a climbing dislocation on the superfluid field, 
we solve the DGPE (Eq.~(\ref{dgpeeq1})) 
with the dislocation potential 
now taking the form as in Eq.~(\ref{tpogpe}) and
the initial condition presented above.  
Given that a stationary dislocation line enhances superfluidity\cite{toner} 
in  its vicinity,  
one naively 
expects that the motion of the dislocation line could
`smear' the superfluid field over a larger region. This  
could then perhaps
suppress the effectiveness of the dislocation line
in enhancing superfluidity, as compared to the stationary case. 
In the results presented in Fig.~\ref{fig1gpe}, 
the dislocation line is assumed to move along the positive $y$ direction 
(climb) at two 
 different speeds 
 namely $V_D =
5\times10^{-4}$  and $1.5\times10^{-3}$ in our
dimensionless units. The value $7.5\times10^{-4}$ was also
studied, yielding intermediate results. 
The response of the superfluid field due to this  
dislocation line motion is illustrated in Fig.~\ref{fig1gpe} through a 
plot of $|\bar{\psi}(\bar{x}=0,\bar{y};\bar{t})|$ at different times. 
At any time $\bar{t}$, the dislocation line is displaced
in the positive $y$ direction by a distance  $V_{D}\bar{t}$. 
The top panel of Fig.~\ref{fig1gpe} shows a plot of 
$|\bar{\psi}(\bar{x}=0,\bar{y};\bar{t})|$ at $\bar{t}=$ 0, 6000, and 14000 for 
$V_D = 5\times10^{-4}$. The 
bottom panel shows the same quantity for $V_D = 1.5\times10^{-3}$ at three 
different 
values of $\bar{t}$, $\bar{t}=$ 0, 4000, and 6000. 
For $\bar{t}=6000$ and $V_D = 1.5\times10^{-3}$, (bottom panel) the 
dislocation line has moved a distance 
of magnitude $\sim 9$, while the corresponding maximum distance
in the top panel is $\sim 7$. 
The shift in the superfluid distribution
as a result of dislocation motion at other values of $\bar{t}$ and $V_{D}$ can 
be clearly observed. 
The plot of $|\bar{\psi}(\bar{x}=0,\bar{y};\bar{t})|$ has a maximum at 
a location $\bar{y}\equiv \bar{y}_{max}$. 
At $\bar{t}=0$, $\bar{y}_{max}$ is at $-1.3$. 

\begin{figure}
\includegraphics[angle=-90, width=0.45\textwidth] {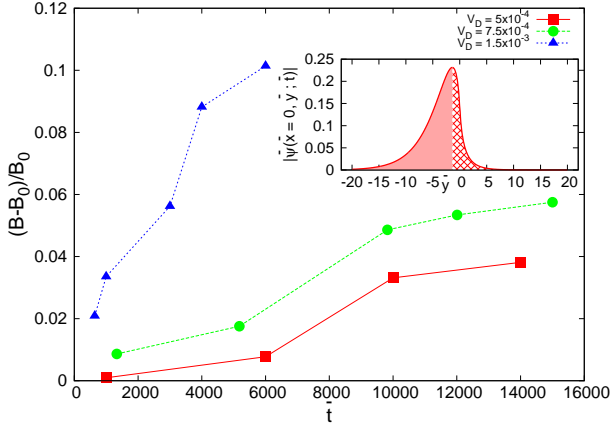} 
\caption{The asymmetry parameter $(B-B_{0})/B_{0}$ (see text)  
during climb motion is shown as a function of time $\bar{t}$ in
the main plot 
 for three different values of $V_D$ as indicated in the legend.
 The inset illustrates the procedure employed to extract $B$ as 
 defined in Eq.~(\ref{asym}). The value for $B_{0}~=~0.538$.} 
\label{fig2gpe}
\end{figure}

As the dislocation line executes climb motion, 
it appears from the figures that the superfluid
distribution becomes more asymmetric in the $y$ direction. 
This increase in asymmetry, with a longer tail towards 
the smaller $y$ region (between negative values of $\bar{y}$ till $\bar{y}_{max}$), might be thought of as if 
some of the  
superfluid amplitude were `left behind' i.e. as the dislocation line moves 
it `smears' the superfluid field over a wider region. 
In order to  
make this more evident, and to quantify it,
we define an asymmetry 
parameter $B$. The $B$ parameter is defined in terms of
the integrated norm of superfluid field in the region 
$\bar{y}<\bar{y}_{max}$ vs in the region $\bar{y}>\bar{y}_{max}$
over the 2D $x-y$ plane, in this way:
\begin{equation}
\label{asym}
B = \frac{\int^{+\infty}_{-\infty}\int^{\bar{y}_{max}}_{-\infty} |\bar{\psi}|^2\,d\bar{x} d\bar{y} - \int^{+\infty}_{-\infty}\int^{+\infty}_{\bar{y}_{max}} |\bar{\psi}|^2\,d\bar{x} d\bar{y}}{\int^{+\infty}_{-\infty}\int^{\bar{y}_{max}}_{-\infty} |\bar{\psi}|^2\,d\bar{x} d\bar{y} + \int^{+\infty}_{-\infty}\int^{+\infty}_{\bar{y}_{max}} |\bar{\psi}|^2\,d\bar{x} d\bar{y}}.
\end{equation}
The procedure implied by this definition is 
illustrated in the inset of Fig.~\ref{fig2gpe}: the contribution to the first 
term 
in the numerator of $B$ in Eq.~(\ref{asym}) is shaded 
in solid color along the plane defined by $\bar{x}=0$. Similarly, 
the contribution to the second term in the numerator is marked by the 
hatched region. 
Due to the motion of the dislocation line in the positive $y$ direction,
if the superfluid field is `left behind', the distribution of $|\bar{\psi}|$ in 
the region $\bar{y}<\bar{y}_{max}$ will increase 
while decreasing in the region $\bar{y}>\bar{y}_{max}$. 
The time dependent 
parameter $B$ can therefore be used to measure the asymmetry in 
the distribution of the superfluid field due to dislocation movement. 
The equilibrium solution shown in Fig.~\ref{fig1gpe}
i.e. $|\bar{\psi}(\bar{x}=0,\bar{y};\bar{t}=0)|$
is asymmetric along the $y$ direction and has a non zero value of the asymmetry 
parameter, 
$B(\bar{t}=0)=B_{0}$. To study the change in asymmetry due to climb motion, we 
look at $(B-B_{0})/B_{0}$
as a function of $\bar{t}$.   
A plot of $(B-B_{0})/B_{0}$ vs $\bar{t}$ is presented in the main part of 
Fig.~\ref{fig2gpe}. Results for the three different values of $V_D$ mentioned
above are given.
The asymmetry of the superfluid field near the dislocation line and along the 
direction of motion 
increases due to climb. 
It is also seen that the rate of increment of
parameter $B$ slows as the dislocation line evolves for longer times. 
For higher climb velocities, more of the superfluid
field tends to be `left behind': 
faster
moving dislocations leave behind more of the superfluid
field. Examination of 
the  wavefunction at $\bar{y}=\bar{y}_{max}$ i.e. 
$|\bar{\psi}(\bar{x},\bar{y}=\bar{y}_{max};\bar{t})|$
shows that climb has no effect on the 
wavefunction shape in the $x$ direction. No change in superfluid field
distribution is observed perpendicular to the direction of climb motion. 

\begin{figure}
\includegraphics[width=0.45\textwidth] {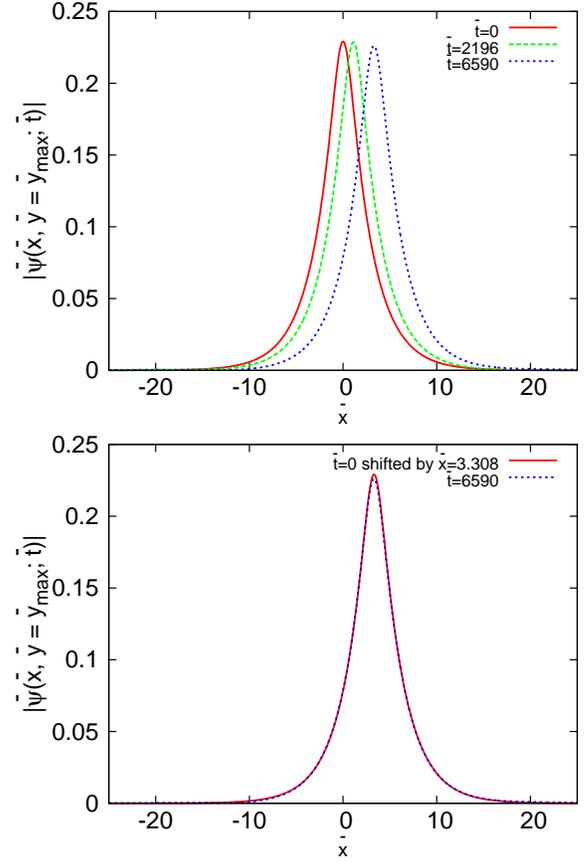} 
\caption{In the top panel, the absolute
value  of the wave function 
$|\bar{\psi}(\bar{x},\bar{y}=\bar{y}_{max};\bar{t})|$ is shown for 
$\bar{t} = 0$, $2196$ 
and $6590$ during glide. $V_D = 5\times10^{-4}$ is used. In the bottom panel,
the equilibrium $|\bar{\psi}(\bar{x}=\bar{x}_{max},\bar{y};\bar{t}=0)|$ is offset by
$\bar{x} = 3.308$ along the positive $x$ direction in order to compare it to
$|\bar{\psi}(\bar{x},\bar{y}=\bar{y}_{max};\bar{t}=6590)|$.}
\label{fig3gpe}  
\end{figure}

\begin{figure}
\begin{turn}{-90}
\includegraphics[width=0.35\textwidth] {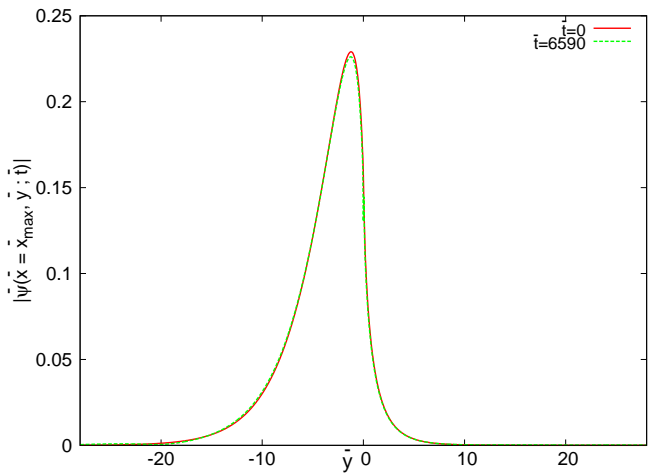} 
\end{turn}
\caption{Plot of $|\bar{\psi}(\bar{x}=\bar{x}_{max},\bar{y};\bar{t})|$ at $\bar{t} = 0$ and $\bar{t} = 6590$
during glide for $V_D = 5\times10^{-4}$. 
No change in the shape $|\bar{\psi}|$ along $\bar{x}=\bar{x}_{max}$ is observed. }
\label{fig4gpe}
\end{figure}

Next, we consider 
glide motion of the dislocation line and the response of the superfluid field
to it. 
We solve the DGPE 
with the dislocation potential 
given in Eq.~(\ref{tpogpe1}) for glide along the positive $x$ direction. 
The top panel of Fig.~\ref{fig3gpe} shows $|\bar{\psi}(\bar{x},\bar{y}=\bar{y}_{max};\bar{t})|$ at
$\bar{t} =$ 0, 2196 and 6590 for $V_D = 5\times10^{-4}$. 
The superfluid field is carried along with the dislocation line.  
The maximum of $|\bar{\psi}|$ at $\bar{y}=\bar{y}_{max}$ shifts from $\bar{x}=0$ to 
a value corresponding to $V_{D}\bar{t}$ referred to as $\bar{x}_{max}$. After $\bar{t}=6590$, the 
maximum of $|\bar{\psi}|$ along $y_{max}$ is expected to shift by $V_{D}\bar{t} = 3.3$, matching the 
simulation results. 
We look again for evidence of asymmetry developing in the superfluid distribution
due to glide motion. 
Glide evolution of $|\bar{\psi}|$ along 
the $x$ direction does not alter its symmetry
or its shape at all, 
as is evident from the bottom panel of Fig.~\ref{fig3gpe}. 
The $\bar{y}=\bar{y}_{max}$ cross section of the 
equilibrium solution $(\bar{t} = 0)$ is shifted
by $\bar{x} = 3.308$ in order to compare it to the time evolved
wave function $|\bar{\psi}(\bar{x},\bar{y}=\bar{y}_{max};\bar{t}=6590)|$. 
No change in the symmetry characteristics for  
$|\bar{\psi}|$ along $\bar{y}=\bar{y}_{max}$ is observed, confirming that 
glide motion 
does not leave behind the superfluid field along the direction of motion.  
Similarly, a plot of $|\bar{\psi}|$ for glide motion 
along the perpendicular direction at  
$\bar{x}=\bar{x}_{max}$, is presented in 
Fig.~\ref{fig4gpe}. We compare there 
$|\bar{\psi}(\bar{x}=\bar{x}_{max},\bar{y};\bar{t})|$ at 
$\bar{t}=0$ and $6590$.  
No change in the shape is observed. 

\begin{figure} 
\includegraphics[width=0.45\textwidth] {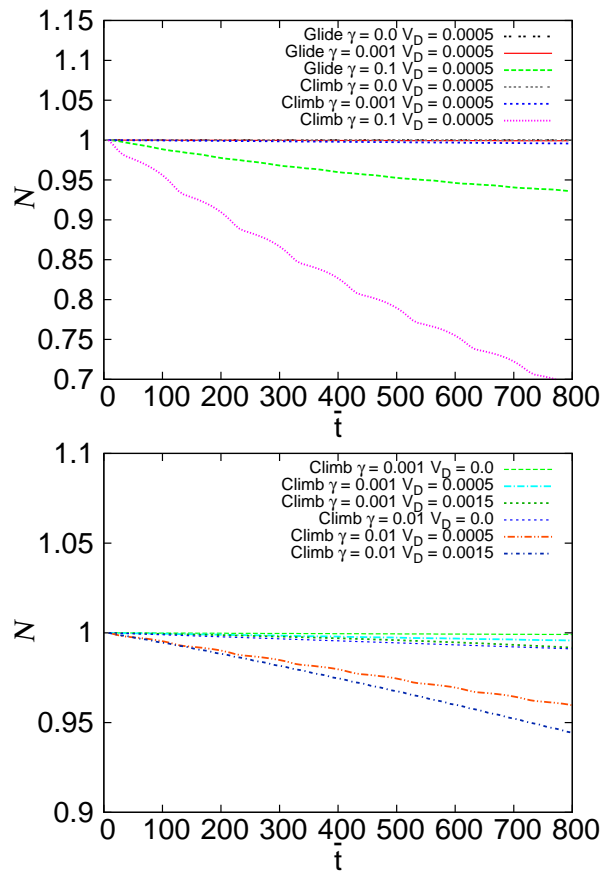} 
\caption{Plots of the total normalization $\mathcal{N}$ vs $\bar{t}$. In the
top panel,  
three different values of $\gamma = 0,~10^{-3}$ and $10^{-1}$ are used for 
both climb and glide motion at a fixed velocity of $V_D~=~5\times10^{-4}$.
In the bottom panel the change in $\mathcal{N}$ at varying velocities 
$V_D = 0,~5\times10^{-4},1.5\times10^{-3}$ for climb at two different values of $\gamma = 10^{-3},~10^{-2}$ is shown.}
\label{fig5gpe}
\end{figure}

We have also studied the 
time dependence of the total normalization of the wavefunction ($\mathcal{N}$
see Eq.~(\ref{normgpe})).
As noted earlier, the damping factor $\gamma$ in the DGPE implies that 
$\mathcal{N}$ is not conserved.  
$\mathcal{N}$ for the superfluid field is observed to decrease for both 
climb and glide motion.
Plots of $\mathcal{N}$ vs
$\bar{t}$ for climb and glide motion at three different values of $\gamma$ are
presented in the top panel of Fig.~\ref{fig5gpe}. 
The decay in $\mathcal{N}$ as a result of glide motion at $V_D = 5\times10^{-4}$
for $\gamma = 10^{-3}$ is too small to be seen and unimportant. An artificially 
larger value of $\gamma = 10^{-1}$ is used 
to amplify any possible decay effect. This results in a $\sim 5\%$ decay 
in $\mathcal{N}$ over a time interval of 800 for glide. 
Climb motion also results in the damping of superfluidity near an edge dislocation line. 
At $\gamma = 10^{-3}$, again, the decay in $\mathcal{N}$ is 
minute. Using a larger value of $\gamma = 10^{-1}$ 
the decay effect is much more visible. Approximately a $30\%$ decay in $\mathcal{N}$ can now be
observed over a time interval of 800. Overall, in comparing climb and glide motion, the decay in 
$\mathcal{N}$ is much more pronounced due to climb. 

The physical origin of the decay in $\mathcal{N}$ can be roughly understood from the following arguments.
Rewriting the DGPE in Eq.~(\ref{dgpeeq}) as 
\begin{equation}
\label{gpen}
i\hbar\frac{\partial\psi}{\partial t} = (1-i\gamma)[H - \mu]\psi,
\end{equation}
where $H = -\frac{\hbar^2}{2m}\nabla_{x,y}^2 + v(x,y) + g|\psi|^2$
it can be seen from Eq.~(\ref{gpeti}) that $H\psi = \tilde{\mu}(t)\psi$. 
By rescaling $t$ 
in the equation above to $t' = (1-i\gamma)t$ 
a solution of the form $\psi = \psi_0 e^{-i\Delta\tilde{\mu}(t)t'}$ is obtained
where $\Delta\tilde{\mu}(t)\equiv\tilde{\mu}(t)-\mu$ is the change in the effective chemical potential.  
This implies that
\begin{equation}
\label{gpen1} 
\psi = \psi_0 e^{-(i/\hbar)(\tilde{\mu}(t)-\mu)t}e^{-(\gamma/\hbar)(\tilde{\mu}(t)-\mu)t},
\end{equation} 
where the damping contribution to $\psi$ can be seen to depend on $\gamma$ and $\tilde{\mu}(t)-\mu$. 
At $\gamma~=~0$ no decay in $\mathcal{N}$ is observed consistent with what is expected from 
Eq.~\ref{gpen1}. 
The quantity $\tilde{\mu}(t)-\mu$ in dimensionless units turns out to be roughly of order $V_D$. 
The 
faster the motion of the dislocation line, the larger the change $\Delta\mu(t)$.

In the bottom panel of Fig.~\ref{fig5gpe} we study the effect of climb velocity (we have already seen that glide is less effective) 
on the damping of the superfluid field 
near the dislocation line at $\gamma~=~10^{-3}$ and an intermediate value  $\gamma~=~10^{-2}$. 
At $V_D~=~0$ and $\gamma~=~10^{-3}$, the decay in $\mathcal{N}$ is too small to be seen. 
For larger values of $V_D$, the decay in the superfluid field is  larger. 
This can be understood by considering that motion of the dislocation line introduces 
excitations into the system thereby raising the $\tilde{\mu}(t)$.
The excitations are responsible for the decay in the superfluid field amplitude.
We see then that for realistic values of $\gamma$ the effect on the overall
normalization is quite small for either climb or glide motion. 

\section{Conclusion}
\label{conclusion}
In this paper we have studied, at the microscopic level,
 the dynamic response of
a superfluid field  associated with an edge dislocation line 
which is moving at a constant speed ${V}_D$. 
Both types of dislocation motion (climb and glide)
are analyzed, for several values of $V_D$.
We use the dissipative Gross-Pitaevskii equation:   
damping of the superfluid field due to dislocation motion
is taken into account via a damping factor $\gamma$, 
as seen in Eq.~(\ref{dgpeeq1}). 
We use a split-step Crank-Nicolson method~\cite{adhikari}  to solve
the DGPE. 
The results provide insight into how the dislocation motion 
influences the evolution of the
superfluid distribution and its damping.  

We determine our initial
conditions by  solving  the equilibrium GPE for the superfluid
field minimally coupled to a stationary dislocation line. 
This solution shows the enhancement of superfluidity 
near the dislocation line -  
the dislocation strain potential acts as a trap for the superfluid field. 
Hence the equilibrium 
wave function $|\bar{\psi}(\bar{x},\bar{y};\bar{t}=0)|$ reflects 
the symmetry characteristics of the strain potential: 
it is symmetric in  $x$ (the direction
of the Burgers vector) at fixed $y$,
e.g. $\bar{y}=\bar{y}_{max}$ and asymmetric in $y$
at fixed $x$, e.g  $\bar{x}=0$. We then solve for the time dependent field
when the dislocation moves. We find that
the superfluid field response to climb shows evidence of superfluidity 
being `left behind' the moving dislocation:  the superfluid
distribution becomes increasingly asymmetric along the direction of
climb. We introduce (see Eq.~(\ref{asym}) and Fig.~\ref{fig2gpe})
an asymmetry parameter $B$  to quantify how the
superfluid field is being `left behind.' 
The parameter $B$ increases as a function of time: it rises quickly at shorter times
and flattens as the dislocation line evolves over a longer time. 
This is consistent with earlier proposals~\cite{balibar_nature}, 
that fluctuations associated with a dislocation line can
be  expected to suppress the associated 
superfluid field, possibly by smearing it over a wider region. 
The magnitude of the asymmetry parameter $B$ 
increases as the dislocation line moves faster. 
At higher speeds, the superfluid field is smeared or left behind over a larger area. 
Therefore, a sudden change in the position of the dislocation 
line makes it more difficult for the superfluid field to be trapped in the dislocation
potential. 
For glide motion, we have also analyzed
the symmetry characteristics of the wave function
at  fixed $\bar{x}=\bar{x}_{max}$ and at fixed $\bar{y}=\bar{y}_{max}$. 
In this case, as opposed to what occurs for climb motion, 
no change in the superfluid distribution 
symmetry characteristics are noted along the glide direction. 
We therefore identify a clear difference in the superfluid response to climb  as compared to glide motion: 
while climb tends to leave behind the initially trapped superfluid field, glide 
movement is quite effective in `carrying along' 
 the superfluid field.  

Both dislocation climb and glide lead 
to a small decay in the superfluid wavefunction normalization ($\mathcal{N}$) for the physical 
value of $\gamma = 10^{-3}$ considered in this study. 
Using a larger 
value of $\gamma = 10^{-1}$, a much larger decay effect can be observed. 
As the parameter $\gamma$ takes into account the energy 
dissipated by the dislocation motion, larger values of $\gamma$ must
indeed, as shown, lead to 
more damping of the associated superfluid field. 
By studying the fluctuations in the amplitude of the superfluid field, a non-hydrodynamic variable, within the DGPE formalism, 
we observe  similar trends in the asymmetric distribution of the superfluid field (as quantified by the parameter $B$)
and the decay in superfluid wavefunction normalization. Faster motion
of the dislocation line leads to both larger decay of the superfluid field and increased asymmetry in its spatial distribution. 
The coupling between the damping parameter $\gamma$ and $V_D$ is clearly elucidated. 

In summary, we have studied the effects of dislocation motion 
on an associated superfluid field. 
During glide, no change in the superfluid field asymmetry characteristics 
is observed. However, climb motion leads to the superfluid field 
being asymmetrically smeared near the dislocation line.
The asymmetry induced in the superfluid distribution due to climb is the most
prominent physical effect observed in this study. 
The implication of dislocation motion in terms of the decay of the superfluid field is also discussed. 

\end{document}